\documentclass[runningheads]{svmult}

\usepackage{makeidx}   
\usepackage{graphicx}  
\usepackage{subeqnar}  
\usepackage{multicol}  
\usepackage{physprbb}  
\makeindex             

\def\HII    {H{\sc {II}}}
\def\etal   {{et~al.\/}}

\begin{document}
\title*{Finding Signatures of the Youngest Starbursts}
\toctitle{Finding Signatures of the Youngest Starbursts}
%
%
\titlerunning{Signatures of the Youngest Starbursts}
%
\author{Henry A. Kobulnicky\inst{1}
\and Kelsey E. Johnson\inst{2} }

\authorrunning{Kobulnicky}
%
%
\institute{University of Wisconsin, Department of Astronomy, 475 N. Charter St.,
Madison, WI 53706
\and JILA, University of Colorado, Boulder, CO 80309}

\maketitle              
\begin{abstract}

Embedded massive starclusters have recently been identified in several
nearby galaxies by means of the radio-wave thermal bremsstrahlung
emission from their surrounding \HII\ regions.   Energy requirements
imply that these optically-obscured starclusters contain 500-1000
O-type stars, making them similar to the ``super starclusters''
observed in many dwarf starbursts and mergers.  Based on their high
free-free optical depth and visual extinctions of $A_V\gg10$ mag.,
these massive ``ultra-dense'' \HII\ regions  (UDHIIs) are distinct
signatures of the youngest, most compact super starclusters.  UDHII
regions may represent the earliest stages of globular cluster
formation.    We review the properties of presently-known UDHIIs, and
we outline a pictoral evolutionary taxonomy for massive cluster
formation which is analogous to the more familiar evolutionary sequence
for individual stars.

\end{abstract}

\noindent{\bf The Youngest Stages of Massive Star Formation } 
\vskip 0.2cm

Massive ($M>10 M_\odot$) stars are observed predominantly in dense
clusters or OB associations.  They are associated with high visual
extinctions and large masses of molecular gas (Brown \etal\ 1999).
Current theories of massive star formation are incomplete.  Some
propose that massive stars form by molecular cloud collapse and
subsequent accretion like low-mass stars, while others  require
that massive stars form via mergers and accretions of stellar-mass
objects in high-density (${\rho}>10^4$ star/pc$^3$) molecular clumps at
the bottom of the cluster gravitational potential (Bonnell, Bate, \&
Zinnecker 1998 and refs therein).  Within massive molecular
star-forming complexes (e.g., W49 in the Milky Way), the youngest
massive stars are seen only indirectly via thermal
bremsstrahlung emission from their surrounding \HII\ regions.  These
ultra-compact \HII\ regions (UC\HII) have sizes of several hundred
A.U., densities of $>10^4$ cm$^{-3}$, emission measures of $>10^7$ pc
cm$^{-6}$, visual extinctions of $A_V>50$ mag., and typically contain
1-2 massive stars (review by Habing \& Israel 1979; Wood \& Churchwell
1989a; Dreher \etal 1984; Wynn-Williams 1971; Becklin, Neugebauer, \&
Wynn-Williams, 1973).  The lifetime of UC\HII\ regions is
$\sim$10\%-15\% of the lifetime of an O star ($\sim$500,000 yr) based
on the observations that 10\%-15\% of Galactic O stars lie obscured
within dense molecular clouds (Wood \& Churchwell 1989b).

Given this picture of the earliest phases of single massive stars, we
might expect that the earliest phases of massive cluster formation
could be modeled as a collection of several hundred UC\HII\ regions.  
Identifying such objects is problematic.  Many nearby galaxies contain
young, massive, blue star clusters with typical ages of several Myr (see
contributions by Whitmore and others in this volume), but these have
been discovered predominantly at optical wavelengths, an approach which
biases the current census of massive starclusters toward the
least-obscured, and, therefore, older, more evolved examples (ages
$\geq$2-3 Myr).  Understanding of the formation and evolution of massive
starclusters requires identifying objects in their formative
proto-cluster stages ($\leq$ 1 Myr).  These stages will be
characterized by enormous visual extinctions ($A_V>50$ mag) which
mandate the use of radio--IR techniques to uncover the
physics of cluster formation.  Recent radio-wave studies have
pinpointed the likely precursors of super starclusters still embedded
in their molecular birthplaces.
\vskip 0.2cm

\noindent{\bf A Case Study of the Blue Compact Galaxy Henize 2-10}
\vskip 0.2cm

The blue compact galaxy Henize 2-10 contains five radio sources with no
optical counterparts and constant flux density over 10 year time
baselines (Kobulnicky \& Johnson 1999.  Figure~1 (upper left) shows a
Hubble Space Telescope F555W  broadband image of the central 300 pc
region along with a VLA 2 cm map (lower left) and an HST H$\alpha$
image.  The radio sources have no optical counterparts in the F555W
image, but  there is a good correlation between the radio map and the
H$\alpha$ morphologies, suggesting that the source of the radio
emission is related to the ionized gas.  The radio sources have
inverted spectral indices between 2~cm and 6~cm
($S_\nu\propto\nu^{+0.5\pm0.3}$), consistent with an optically-thick
thermal bremsstrahlung origin (upper right).

We model the spectral shape and luminosity of the observed sources as
spheres of uniform-density plasma with an electron temperature of 6000
K.   H~II regions with radii between 3 pc and 8 pc, densities of 5000
$cm^{-3}$, and 500-1000 ionizing O7V stars are most consistent with the
data.  These high densities imply an overpressure compared to typical
warm ionized medium pressures.  Such \HII\ regions should expand and
become undetectable in the thermal radio continuum on timescales of
500,000 yr.  Thus, is seems likely that the ages of these \HII\ regions
are very small, consistent with their heavily-obscured nature.  For a
typical Salpeter IMF extending from 0.5 to 100 M$_\odot$, the clusters
contain 6$\times10^5$ total stars, implying peak stellar densities of 5000
$pc^{-3}$.

Several other galaxies harbor UD{\HII}s discovered by their peculiar
radio spectral index and high brightness temperature: one 
in NGC 5253 (Turner, Ho, \& Beck 1998; Turner, Beck, \& Ho 2000);  six
in NGC 2146 (Tarchi \etal\ 2000); 1-2 in NGC 4214 \& Tololo
35 (Beck \etal\ 2000).  Radio recombination line studies of
suggest the presence of UD{\HII}s in NGC 3628 and IC 694
(Zhao \etal\ 1997).

\begin{figure}[b]
\begin{center}
\includegraphics[width=0.7\textwidth,angle=0]{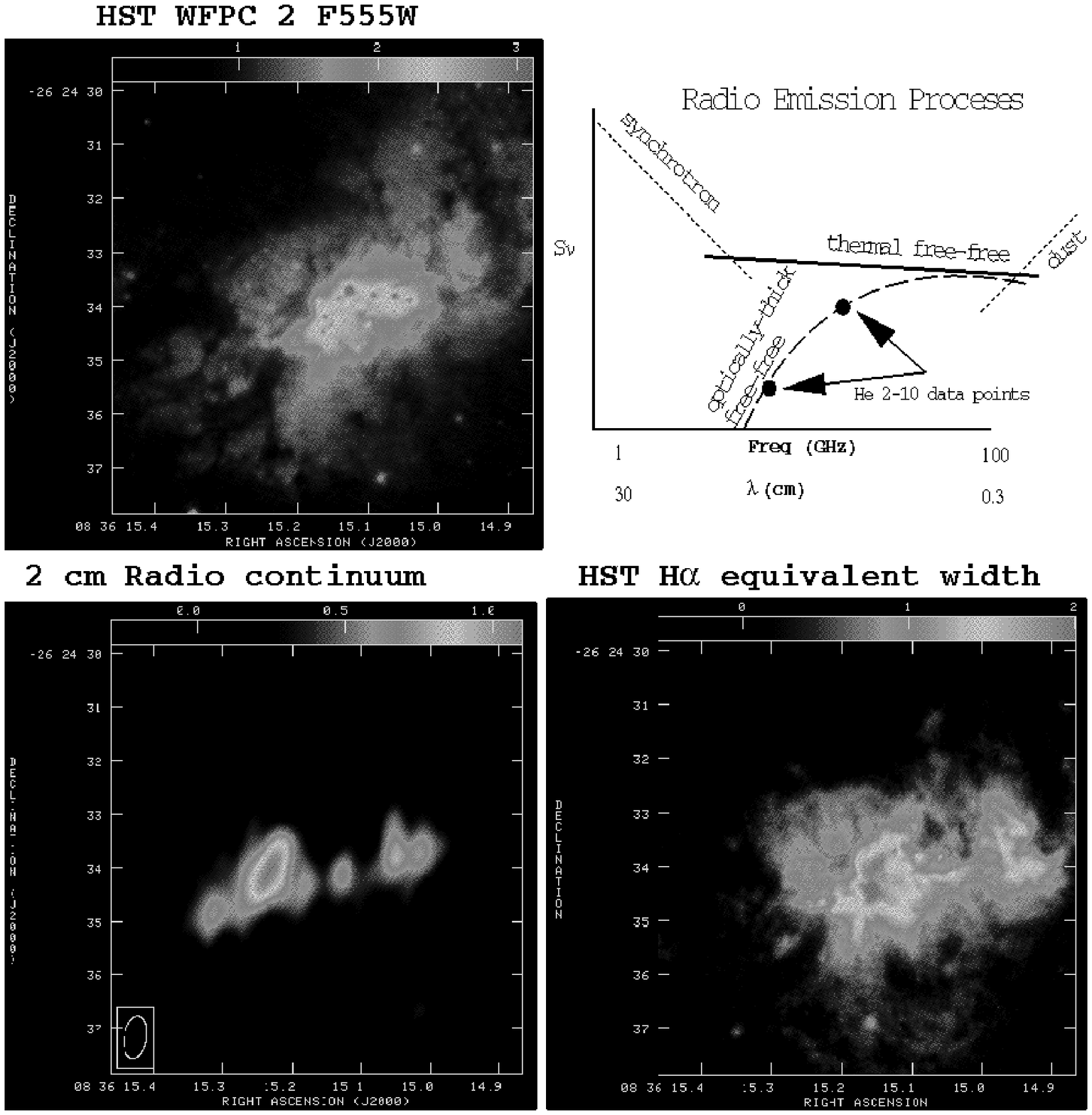}
\end{center}
\caption[]{Images of Henize 2-10 in optical HST F555W (top left), 2 cm radio
continuum (lower left) and H$\alpha$ equivalent width (lower right).  
The schematic at the upper right illustrates the frequency
dependence of non-thermal (synchrotron emission), thermal bremsstrahlung,
and optically-thick thermal bremsstrahlung emission mechanisms.  }
\label{images}
\end{figure}

\begin{figure}[b]
\begin{center}
\includegraphics[width=.4\textwidth,angle=-90]{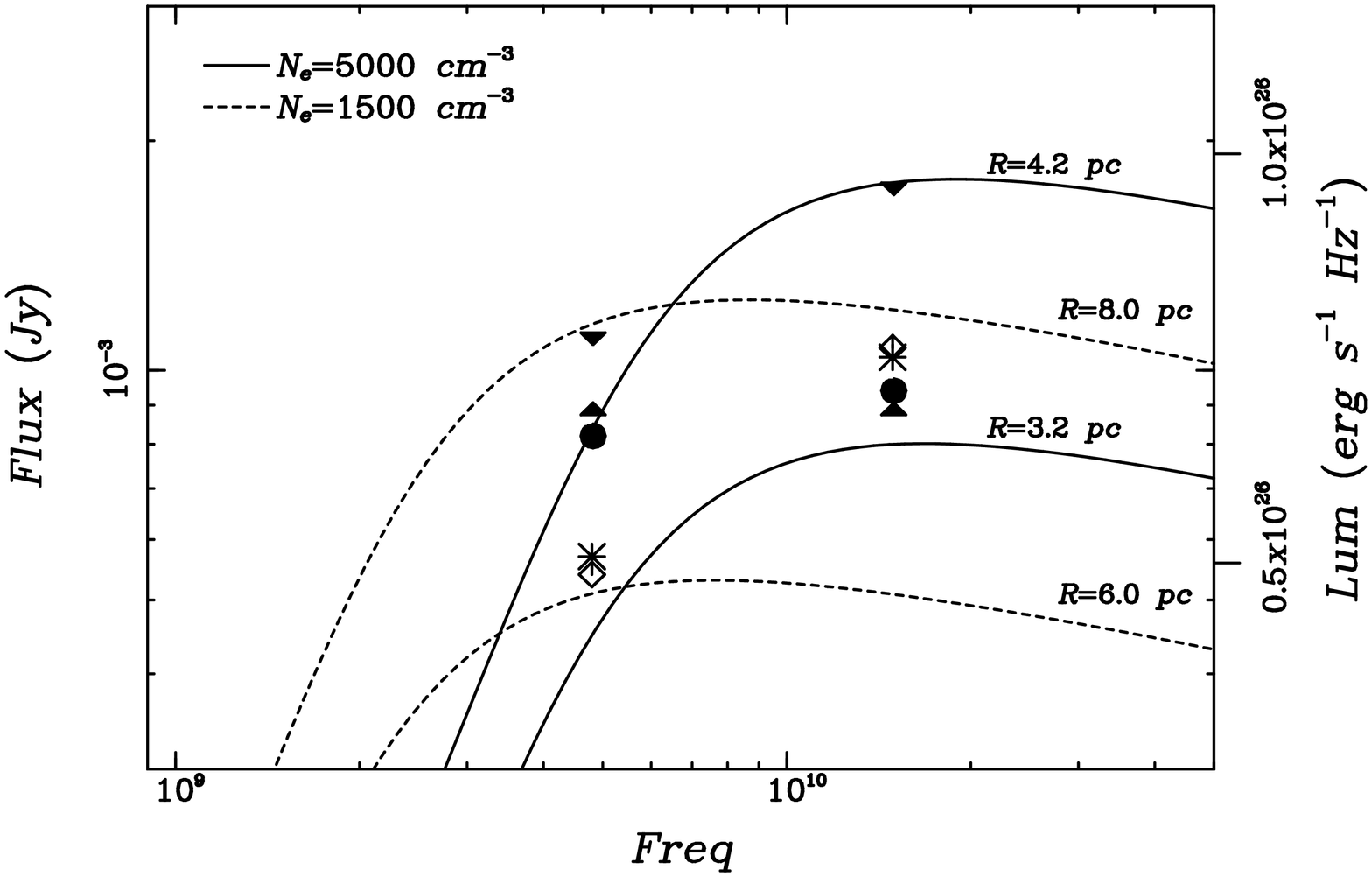}
\end{center}
\caption[]{VLA 6~cm (4.8~GHz) and 2~cm (14.9 GHz) fluxes and luminosities
for the five radio knots in Henize~2-10.  A different symbol represents
data for each knot.  Solid and dotted lines represent
thermal bremsstrahlung plasmas modeled as
spheres of radius, $R$, electron temperature $T_e$=6000~K, and mean
electron densities of 1500 $cm^{-3}$ and 5000  $cm^{-3}$ respectively.
 }
\label{models}
\end{figure}

\vskip 0.2cm

\noindent{\bf Towards An Evolutionary Taxonomy for Massive Starclusters  }
\medskip

\indent {\it Low Mass Stars:\ } The evolutionary sequence of low-mass star
formation  has been outlined by Lada (1987) and Andr\'e, Ward-Thompson \&
Barsony (1993).  Low-mass stars begin as prestellar molecular cores,
and evolve through stages (Class 0 through Class III as depicted in
Figure~3) characterized by an increasing faction of infrared emission
from the central star and a decreasing fraction of submillimeter
emission from dust in the accretion disk.

{\it Massive Stars:\ } For massive stars, no clearly defined
evolutionary sequence has yet been agreed upon.  In Figure~3 (center
column) we attempted to sketch an evolutionary sequence based on
current literature.  At the earliest times, massive stars begin a a
collection of dense molecular cores {\bf(1)}.   Because there are
theoretical problems with forming stars more massive than
$\sim10~M_\odot$ in the collapse of a single molecular core (Bonnell
\etal\ 1998 and refs therein), mergers among nearly-formed proto-stars
are invoked to produce the most massive objects {\bf(2)}.  
Once accretion or merging terminates and the massive
star is formed, high-energy photons begin to ionize the surrounding
molecular gas, producing an UC\HII\ region {\bf(3)}.  UC\HII\ regions
have sizes $\leq$0.1 pc, and often exhibit  cometary morphologies.
The massive star hidden within an UC\HII\ region
emerges from its natal molecular cloud through the combined actions of
its stellar wind, ionizing radiation, and space velocity {\bf(4)}.
Massive stars spend the latter 80\% of their lifetimes visible as
luminous O stars {\bf(5)} (see review by Churchwell 1990).

{\it Massive Star Clusters:\ } In the case of massive starclusters
(Figure~3, right column), our theoretical and observational
understanding of their evolutionary phases is even less secure.
Because bound massive stellar clusters contain $>few\times10^5$
$M_\odot$ of stars in a region just a few pc in size (Ho \& Filippenko
1996; O'Connell, Gallagher, \& Hunter 1994) they must begin with the
collapse and fragmentation of exceptionally compact and massive
molecular structures exceeding $10^7$ $M_\odot$.  We term these
``Massive Molecular Aggregates'' in Figure~3 {\bf (1)}.  The star
formation efficiencies in massive clusters must exceed the typical
values of $\sim$0.5\% in Galactic SF regions (Lada 1987) and 2\%--5\%
in M33 OB associations (Wilson \& Matthews 1995) in order to form the
required mass of stars from typical $10^4$--$10^6$ $M_\odot$ Giant
Molecular Clouds.  Star formation efficiencies exceeding 50\% appear
reasonable in some young clusters (Arp 220---Anantharamaiah
\etal\ 2000; young Galactic clusters---Lada, Evans, \& Falgarone 1997),
and may also help increase the binding energy of the resultant cluster
since there is less residual gas to be swept out of the gravitational
potential (Adams 2000;  Goodwin 1997).  At the center of the
gravitational potential, an aggregate of thousands of warm, massive
molecular cores in a space just a few pc across provides the
environment for the formation of massive stars through mergers and
accretion (Bonnell \etal\ 1998).  Such objects should be detectable as
a collection of extremely compact subillimeter sources, so we term
these ``Massive Submillimeter Aggregates'' {\bf (2)}.  These first two
stages are somewhat speculative schematics predicated on extrapolation
of massive molecular complexes in the Milky Way where giant molecular
clouds range only from 10$^4$-10$^6$ $M_\odot$ and have sizes from
50-100 pc (Sanders, Scoville, \& Solomon 1985).  To our knowledge, {\it
there have not yet been observed any molecular structures which are
sufficiently massive and compact to be identified as the likely
predecessors of super starclusters.}  The newly-identified class of
ultra dense \HII\ regions (UD{\HII}s) are the massive analogs of the
more familiar single-star UC{\HII}s, and they represent a transition
phase {\bf (3)} between the warm molecular and massive submillimeter
aggregates and the familiar UV-bright starclusters  {\bf (4, 5)}.

\begin{figure}[b]
\begin{center}
\includegraphics[width=.99\textwidth]{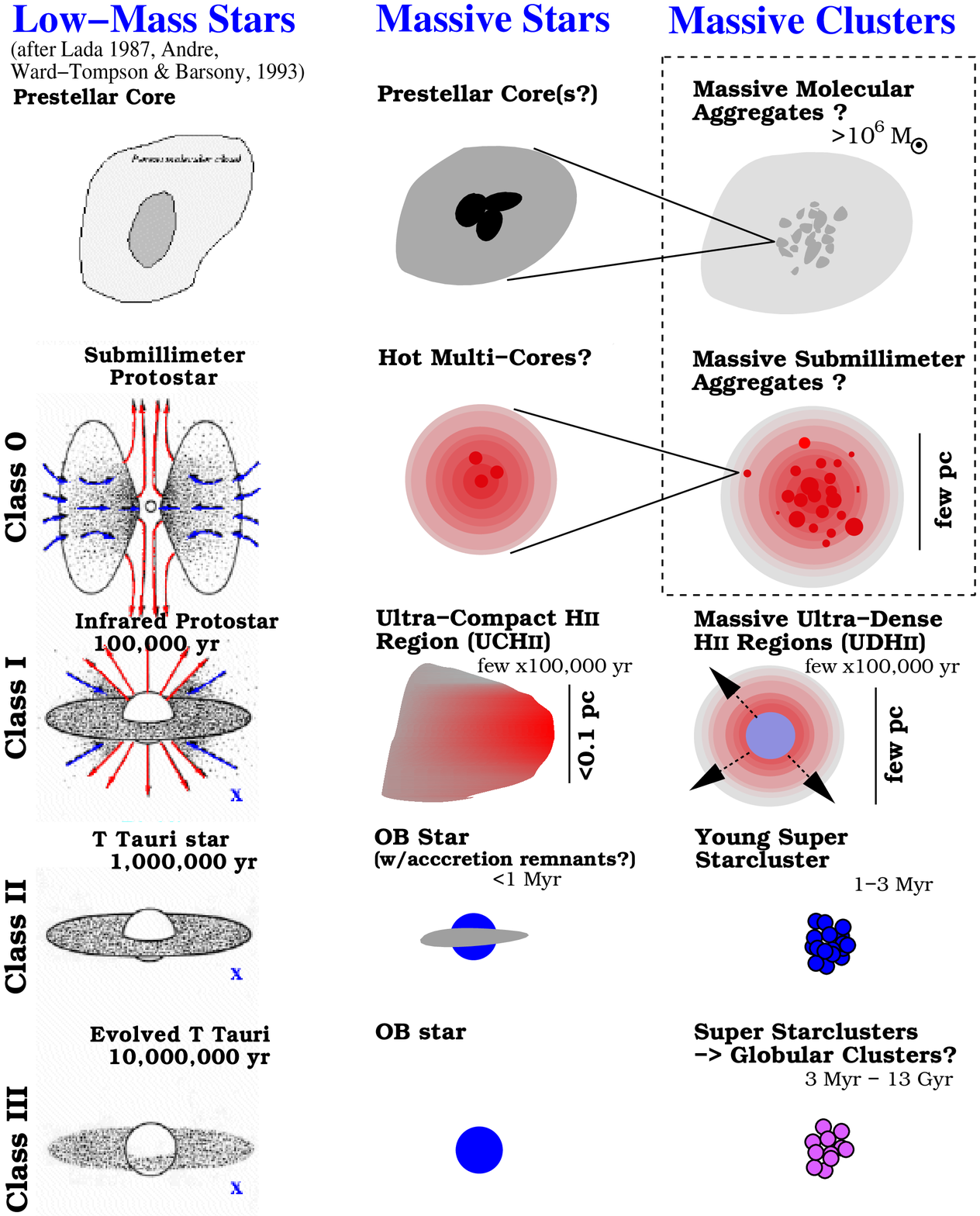}
\end{center}
\caption[]{A tentative, and somewhat speculative,
schematic illustrating the stages of low-mass star 
formation (left, from Lada 1987; Andre, Ward-Thompson, \& Barsony 1989),
massive star formation (center) and massive cluster formation (right).
Ultra-dense HII regions represent the youngest phase of
massive cluster formation yet identified.   Dashed lines surround
phases of massive cluster formation which are speculative 
and have no identified examples.}
\label{stages}
\end{figure}

Unusually compact, massive ($10^7$ $M_\odot$) gravitationally-bound
concentrations of molecular gas and dust on size scales of 2-6 pc could
signify the genesis sites of super starclusters.  However, these early
phases have not yet been specifically identified.  Wilson \etal\ (2000)
report the discovery of massive molecular resevoirs, termed ``super
giant molecular complexes'', in the Antennae merger system (NGC
4038/39) which could be the source material for the ``Massive Molecular
Aggregates''.  Given the abundance of massive
extragalactic starclusters, their predecessors, the ``Massive Molecular
Aggregates'' and the ``Massive Submillimeter Aggregates'', should be
detectable in extragalactic sources with current millimeter and
upcoming sub-millimeter arrays.  Identification of these precursors to
massive starclusters is the first step toward understanding the
dynamical conditions that produce OB associations, super starclusters,
and the venerable globular cluster systems.

\vskip 0.2cm

{\it Acknowledgments:}  We are grateful for conversations with Ed
Churchwell, Peter Conti, Jay Gallagher, and Jonathan Tan which have
helped to clarify this presentation.

%

\end{document}